 \documentclass[prl,twocolumn,showpacs,amsmath,amssymb,superscriptaddress]{revtex4}
 \usepackage{graphicx}% Include figure files
\usepackage{dcolumn}% Align table columns on decimal point
\usepackage[sort&compress]{natbib}
\usepackage{subfigure}
\usepackage{ifpdf}
\usepackage{bm}
\ifpdf
\usepackage[pdftex,
        colorlinks=true,
        pdftitle={},
        pdfauthor={S. Salem-Sugui Jr.},
        pdfsubject={fluctuation in iron-pnictides},
        pdfkeywords={IF, UFRJ, Rio de Janeiro Brazil},
        baseurl={http://www.if.ufrj.br},
        pdfpagemode=UseNone,
        bookmarksopen=true
        pdfoutput=1
        ]{hyperref}
\fi

\begin{document}
\title{Superconducting fluctuations in the reversible magnetization of the iron-pnictide $Ba_{1-x}K_xFe_2As_2$}
\author{S. Salem-Sugui Jr.}
\affiliation{Instituto de Fisica, Universidade Federal do Rio de Janeiro,
21941-972 Rio de Janeiro, RJ, Brazil}
\author{L. Ghivelder}
\affiliation{Instituto de Fisica, Universidade Federal do Rio de Janeiro,
21941-972 Rio de Janeiro, RJ, Brazil}
\author{A. D. Alvarenga}
\affiliation{Instituto Nacional de Metrologia Normaliza\c{c}\~ao e
Qualidade Industrial, 25250-020 Duque de Caxias, RJ, Brazil.}
\author{J.L. Pimentel Jr.}
\affiliation{Instituto de F'sica, Universidade Federal do Rio Grande do Sul, Porto Alegre, RS}
\author{Huiqian Luo}
\affiliation{National Lab for Superconductivity, Institute of Physics and National Lab for Condensed Matter Physics, P. O. Box 603 Beijing, 100190, P. R. China}
\author{ Zhaosheng Wang}
\affiliation{National Lab for Superconductivity, Institute of Physics and National Lab for Condensed Matter Physics, P. O. Box 603 Beijing, 100190, P. R. China}
\author{Hai-Hu Wen}
\affiliation{National Lab for Superconductivity, Institute of Physics and National Lab for Condensed Matter Physics, P. O. Box 603 Beijing, 100190, P. R. China}
\date{\today}
\begin{abstract}
We report on isofield magnetization curves obtained as a function of temperature in two single crystals of $Ba_{1-x}K_xFe_2As_2$ with superconducting transition temperature $T_c$=28K and 32.7 K. Results obtained for fields above 20 kOe show a well-defined rounding effect on the reversible region extending 1-3 K above $T_c(H)$ masking the transition. This rounding appears to be due to three-dimensional critical fluctuations, as the higher field curves obey a well know scaling law for this type of critical fluctuations. We also analyzed the asymptotic behavior of $\sqrt M$vs.T curves in the reversible region which probes the shape of the gap near $T_c(H)$. Results of the analysis suggests that phase fluctuations are important in  $Ba_{1-x}K_xFe_2As_2$  which is consistent with nodes in the gap. 
\end{abstract} 
\maketitle 

\section{Introduction}

Among the iron-pnictide systems found until now \cite{japao,milton},  $Ba_{1-x}K_xFe_2As_2$ is one of the most studied. Experiments on this system includes,  local magnetization \cite{twogaps}, resistivity and bulk magnetization \cite{0808.2392,0807.3137}, specific heat \cite{CvLuo,0812.1188,welp}, magneto-resistance \cite{0807.4488}, NMR\cite{0901.0177}, thermal-conductivity \cite{0811.4668}, ARPES \cite{epl,germany}, electronic-relaxation \cite{chia},  vortex-dynamics \cite{0807.3786}, combined muon-spin rotation, neutron scattering and magnetic force spectroscopy \cite{germany2} among others, but fluctuations studies are still lacking, which motivated this investigation. 

Superconducting fluctuations are important in the vicinity of the transition temperature $T_c$ and  have been extensively studied in all kinds of superconductors \cite{klemm1,klemm2,tinkham}. In the case of layered systems, fluctuation superconductivity is enhanced, and there are robust theories predicting the behavior of many measurable quantities with temperature and magnetic field, in or out of the critical region \cite{klemm1,ger,ullah,tesa,rosenstein,kwon,beck,jesus}.  Comparison of experimental results with these theories helps to understand the nature of the fluctuations, its dimensionality \cite{klemm2,rosenstein,nb,said,paulo} and give additional insight about the pairing mechanism symmetry of the studied system \cite{kwon,alvarenga}. 

In this work we study the effect of superconducting fluctuations in the region of reversible magnetization of the new iron-pnictide system  $Ba_{1-x}K_xFe_2As_2$ \cite{bulk,crystal}, which is a fully gaped layered superconductor  \cite{CvLuo}.  This system is a hole doped oxygen-free iron-pnictide superconductor with $T_c$ ranging from 24K to 38 K depending on doping \cite{bulk,crystal}. Similar to the high-$T_c$ cuprates the  $Ba_{1-x}K_xFe_2As_2$ system presents a "bell-shaped" phase diagram of $T_c$ against  doping \cite{0807.3950,norman}, but the system has a low anisotropy  with an anisotropy factor $\gamma$  between 2-3 \cite{0808.2392, 0807.4488} and a s-wave multiband pairing symmetry \cite{twogaps,epl}. A nodeless s-wave gap is claimed in Ref.\onlinecite{epl}, but Ref.\onlinecite{0811.4668} show  evidence for nodes in the gap, and a recent theoretical work shows that a multiband-s-wave gap can have nodes \cite{nodes2}. The multiband s-wave pairing mechanism in iron-pnictides is still on debate, whether it is a conventional s-wave \cite{stanevPRB78,cvetepl} or unconventional with nodes \cite{mazin,nodes2}.

We demonstrate in the present study that, contrary to what was expected \cite{welp}, isofield reversible magnetization curves obtained in $Ba_{1-x}K_xFe_2As_2$ for magnetic fields, H,  in the range of 20-50 kOe show a considerably large amplitude fluctuations in the vicinity of $T_c(H)$ with a well-defined rounding effect which masks the transition. This rounding effect is reminiscent of  magnetization curves obtained in high-$T_c$ superconductors \cite{ullah,welp2,tesa,zac,rosenstein}. In the case of cuprate superconductors,  superconducting activity above $T_c$ appears to be related with the so called pseudo-gap \cite{timusk, kpaper} which is not considered in the theories for layered systems \cite{mauro} . Despite some similarities between iron-pnictides and cuprates in what concerns for instance the existence of magnetic fluctuations and magnetic order (AF anomaly) in the non-superconductor precursors \cite{norman}, most likely there is no such pseudo-gap phase in iron -pnictides \cite{chennature} . Furthermore the study of  fluctuations effects in this iron-pnictide system is of particular interest, allowing further comparison with newly developed theories.

We also address the existence of phase fluctuations in  $Ba_{1-x}K_xFe_2As_2$ by analyzing the asymptotic behavior of $\sqrt{M}$ vs T curves in the reversible region. The existence o nodes in the gap may enhance phase fluctuations effects in the vicinity of $T_c$ \cite{kwon} changing the shape (the asymptotic behavior) of the gap in this region. The quantity $\sqrt{M}$ near $T_c$ is directly proportional to the amplitude of the order parameter \cite{abrikosov}  and $\sqrt{M}$ vs T curves can probe the shape of the gap near $T_c(H)$ \cite{kwon,alvarenga}. Our analysis suggests that phase fluctuations are important near $T_c$ in  $Ba_{1-x}K_xFe_2As_2$.

\section{Experimental} 
Two high quality single crystals of $Ba_{1-x}K_xFe_2As_2$ were investigated, with $T_c$= 32.7  and  28 K, corresponding to a potassium content of x=0.28 and 0.25 respectively, and with masses of approximately 0.05 and 0.3 mg.  Both show a fully developed superconducting transition with width $\Delta T_{c}\simeq 1 K$ for sample with $T_c$= 32.7K and width $\Delta T_{c}\simeq 2 K$ for sample with $T_c$= 28 K. The crystals were grown by a flux-method described elsewhere \cite{crystal}. Similar samples made by the same group were used in various previous investigations \cite{welp,twogaps,0808.2392,0807.3786}. Since anisotropy is low in this system \cite{0808.2392,0807.4488}, the measurements were carried out only for the direction $H\parallel c$-axis. Magnetization data were obtained with a commercial magnetometer, based on a superconducting quantum interference device (SQUID).  The data were obtained after cooling the sample from temperatures well above $T_c$  in zero applied magnetic field (zfc) to a desired temperature below $T_c$ ($T_{min}\simeq 10 K$).  Magnetic fields up to 50~kOe were applied, reaching the desired value without overshoot.  The data were obtained by continuously heating the sample, and collecting magnetization results within fixed increments of temperature, $\delta T\simeq0.1 K$, up to $T_{max}\simeq 80 K$.  We also obtained field-cooled curves (fc), which provide the reversible (equilibrium) magnetization. The background magnetization due to the normal state contribution was determined and removed for each data set by fitting to the form $M_{b}=c(H)/T+a(H)$ in a temperature window well above $T_c$. 
\begin{figure}[t]
\includegraphics[width=\linewidth]{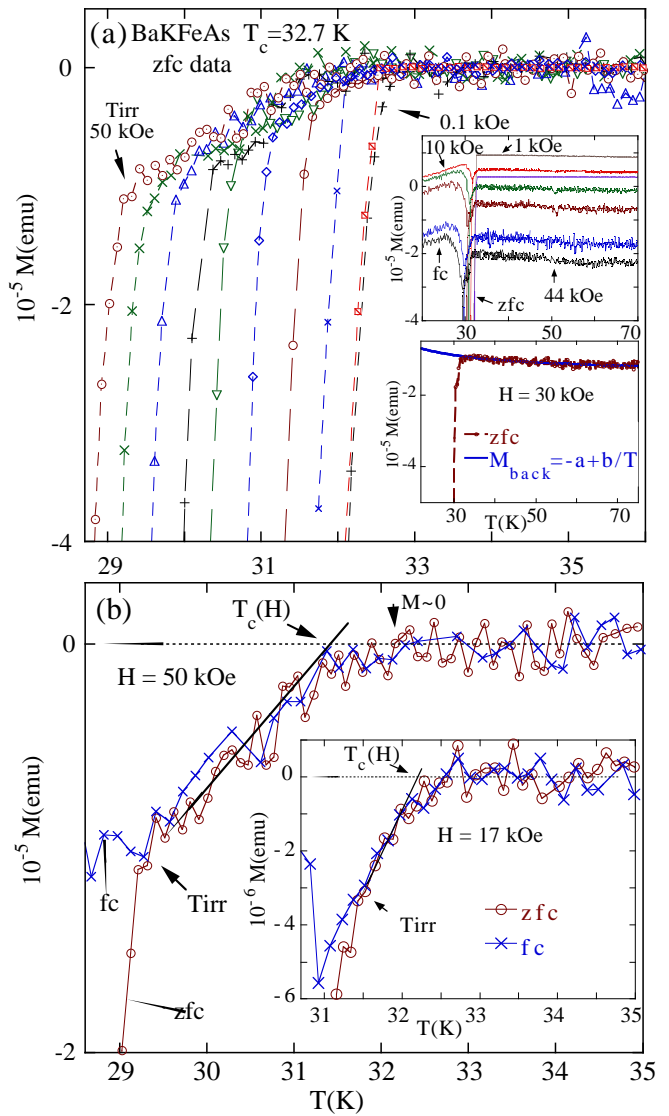}
\caption{Isofield M vs. T  curves for sample with $T_c$=32.7 K. (a) zfc curves for H=0.1, 0.5, 5, 10, 17, 24, 30, 37, 44, 50 kOe. Top inset show zfc-fc curves for (top to bottom) 1, 10, 0.1, 17, 24, 37 and 44 kOe; low inset show detail of the background removal. (b) detail of reversible region for H= 50 kOe. Inset: detail of reversible region for H=17 kOe.}
 \label{fig1}
\end{figure}
\begin{figure}[t]
\includegraphics[width=\linewidth]{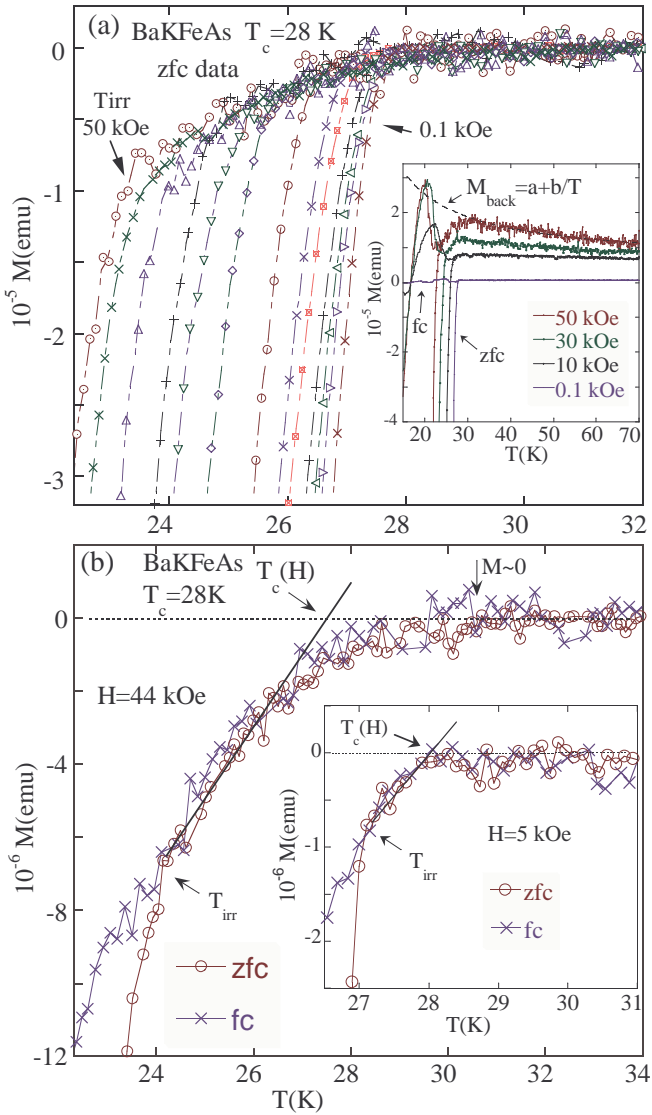}
\caption{Isofield M vs. T curves  for sample with $T_c$=28 K. (a) zfc curves for H=0.1, 03, 0.5, 1, 3, 5, 10, 17, 24, 30, 37, 44, 50 kOe. Insets: zfc-fc curves and background removal. (b) detail of reversible region for H= 50 kOe. Inset: detail of reversible region for H=5 kOe.}
 \label{fig2}
\end{figure}
\begin{figure}[t]
\includegraphics[width=\linewidth]{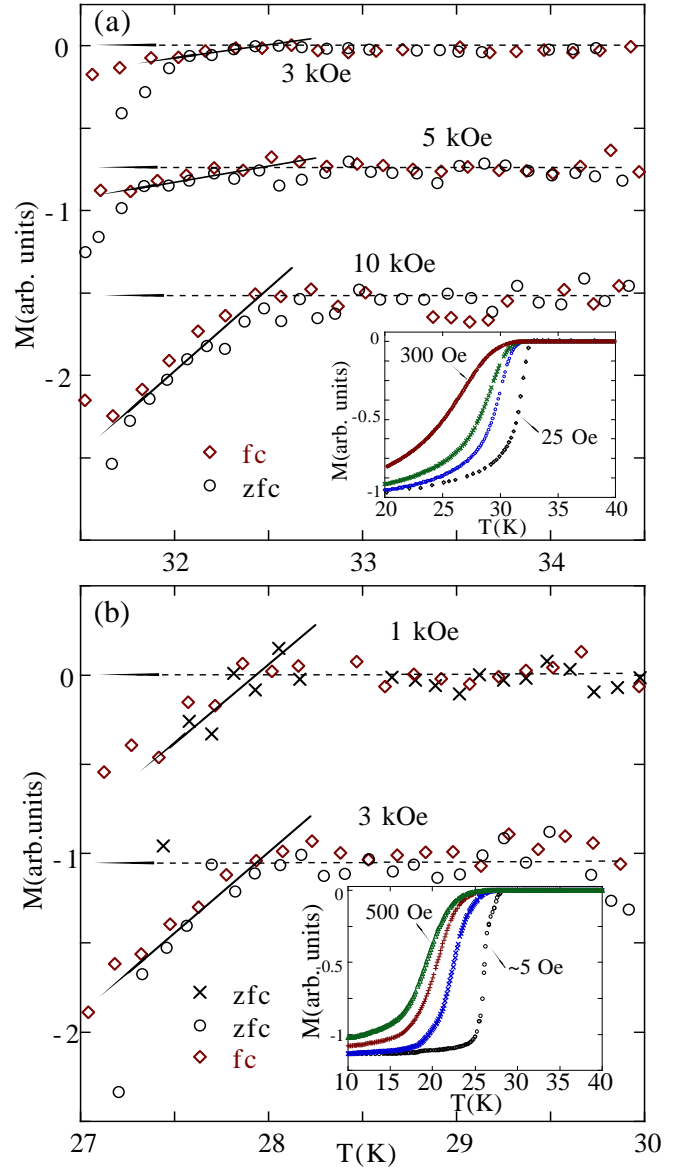}
\caption{Selected low field M vs. T curves showing detail of the reversible region. (a) zfc-fc curves for sample with $T_c$=32.7 K. (b) zfc-fc curves for sample with $T_c$=28 K. The curves were shifted vertically for clarity. Insets: zfc curves for lower fields. The lowest field curve in the inset of Fig.3b was obtained with an ac field of 1 Oe at 1 kHz in the presence of a $\sim 5~Oe$ remnant field of the magnet.}
 \label{fig3}
\end{figure}

\section{Results and discussion}
Figures~1a and ~2a show zfc magnetization curves as obtained for each sample after background removal. Insets of Figs. 1a and 2a show details of selected field-cooled and zero-field-cooled magnetization curves as obtained in a wider temperature range, and illustrates the background removal procedure used in all curves. We observe that the background signal for sample with $T_c$= 32 K is paramagnetic from 100 Oe up to 10 kOe. The paramagnetic signal increases with field up to 1 kOe. Above 1 kOe a field increasing diamagnetic signal suppresses the paramagnetic one, which results in an overall diamagnetic signal for fields higher than 17 kOe. The origin of this effect is unknown. The trend observed for this sample (for the magnetization of the normal state) for fields higher than 10 kOe is similar to the trend observed in $YBaCuO$ and other cuprates. On the other hand, the sample with $T_c$ = 28 K has a paramagnetic background signal which increases with field, as in Pauli paramagnetism. The hump appearing in the field cooled curves just below the irreversible temperature, Tirr, of each curve in the insets of Figs. 1a and 2a was observed for all curves, even for low values of the applied magnetic field.  

The reversible magnetization in each curve of Figs. 1a and 2a corresponds to a region where M approaches zero with a low slope (see arrows marking the position of Tirr for H=50 kOe). This fact is better exemplified in Figs. 1b and 2b where selected curves obtained for  fields higher than 40 kOe show a rather large reversible region with a well-defined rounding effect as M approaches zero. In contrast to that, the insets of Figs. 1b and 2b and Figs. 3a and 3b show curves obtained with lower fields (H$\prec$20 kOe), where the reversible region approaches zero linearly, as expected in a second order phase transition. It is possible to see on the low field curves presented in Figs. 3a and 3b that the linear reversible region clearly defines $T_c(H)$. Insets of Figs. 3a and 3b show details of the zfc low field curves. The results of Figs. 1b and 2b and the respective insets also exemplify the linear extrapolation of the reversible magnetization \cite{abrikosov} used to estimate the mean field value of $T_c(H)$ in each curve. It is possible to visualize on the curves of Fig. 1b and 2b (marked by arrows) that the magnetization reaches zero for temperatures $\sim1 K$ and $\sim3 K$ above $T_c(H)$ respectively. This rounding effect masking the second order phase transition was observed in all curves obtained for fields above 20 kOe, and it is more pronounced in the sample with $T_c$=28 K. This rounding is reminiscent of high-field diamagnetic fluctuations observed in high-$T_c$ superconductors which have been extensively studied \cite{ullah,welp2,tesa,zac,rosenstein}. It should be mentioned that a similar but less pronounced rounding effect have also been observed in Nb with $\kappa = 4$ \cite{nb}, which is a well know BCS superconductor. It is important to mention that, since the rounding effect in the higher field curves extends above $T_c$ which is defined here at the onset of diamagnetic signal,  we discard the possibility of the rounding effect to be influenced by sample inhomogeneity.  Also, the existence of any sample inhomogeneity is expected to produce a clear rounding in the reversible magnetization curves near the transition of low field curves, which is not observed in the reversible magnetization of the low field curves presented in Figs. 3a and 3b. 
\begin{figure}[t]
\includegraphics[width=\linewidth]{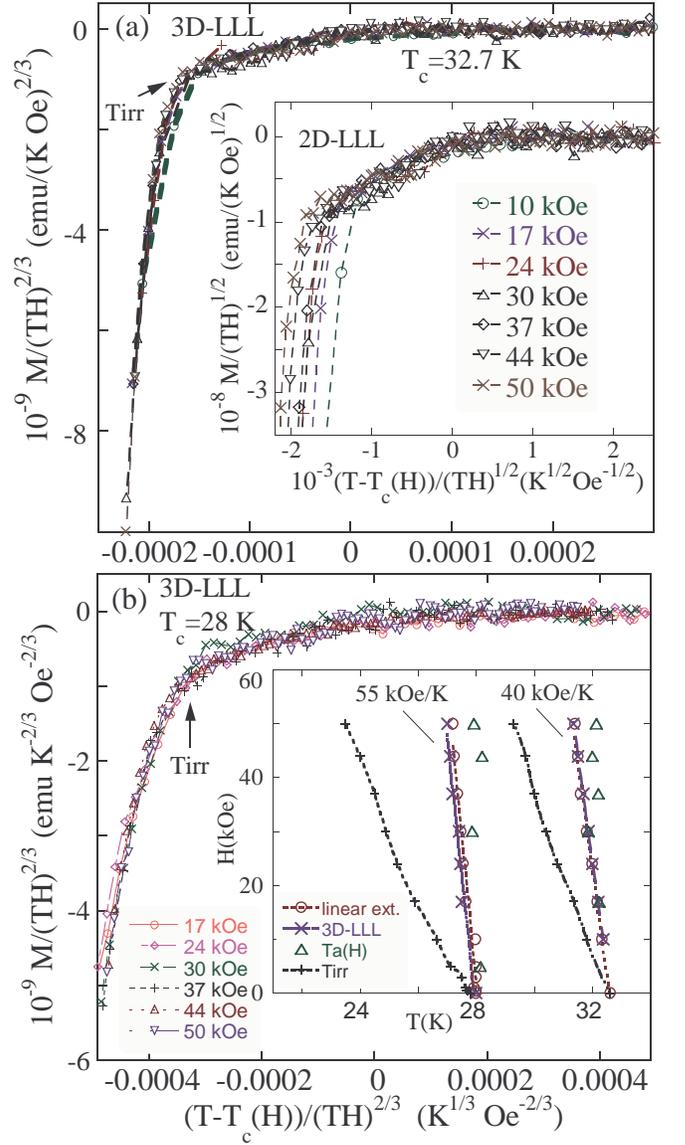}
\caption{Collapse of magnetization curves presented in Figs. 1a and 2a for $H\succ10$kOe after the 3D-LLL scaling. (a) sample with $T_c$=32.7 K. Inset: 2D-LLL scaling of the same data. (b) sample with $T_c$= 28 K. Inset: phase diagram of the studied samples.} 
\label{fig4}
\end{figure}

In the present investigation, we perform a scaling analysis of the magnetization curves using the results obtained in Ref. \onlinecite{rosenstein} for a layered material which consider fluctuations-fluctuations interactions within the Ginzburg-Landau formalism. These authors \cite{rosenstein} obtained expressions for two (2D) and three dimensional (3D) critical fluctuations with the approximation that the magnetic field is high enough for  all carriers to lye in the lowest-Landau-level (LLL). To perform the scaling we replace the temperature $x-axis$ and magnetization $y-axis$ of each curve to the respective scaling forms $(T-T_c(H))/(TH)^{(D-1)/D}$ and $M/(TH)^{(D-1)/D}$ where D is the dimensionality, and plot together all scaled  curves. The only free parameter in the scaling procedure is the mean field temperature $T_c(H)$ which is adjusted for each curve so that all results fall in a single universal curve. 
 
Based on the low anisotropy of the BaKFeAs system we expect that fluctuations are of three dimensional nature. However, since spin-density-waves and superconductivity may coexist in this system for certain values of doping \cite{0807.3950,chia,germany2} and two dimensional magnetic fluctuations have been recently observed above $T_c$ \cite{0901.0177} we also performed a two-dimensional scaling analysis. Figs. 4a and 4b show the results of the scaling for D=3 for both samples. Results for fields lower than 17 kOe fail to collapse in the single curve. The values of $T_c(H)$ obtained from the scaling analysis are in strictly good agreement with values obtained from the linear extrapolation of the reversible magnetization (see Fig. 3). It should be observed that not only the reversible magnetization follows the scaling and collapses into a single curve but also a portion of the  irreversible region lying below the arrow marking the position of Tirr. The inset of Fig. 4a show the results of the two-dimensional scaling using values of $T_c(H)$ obtained from the linear extrapolation procedure. A similar result, not shown, was obtained for the other sample. Even a simple visual comparison between Fig. 4a and its inset unambiguously confirms that the fluctuations are of three dimensional nature. Although it is possible to obtain a good collapsing curve with the two-dimensional scaling, the resulting  $T_c(H)$ values are more than 1 K  lower than the values obtained from the linear extrapolation which is physically meaningless.  

To better quantify the dimensionality of the BaKFeAs system, we estimate the
value of the parameter \cite{ger} $r=8(m/M)[\varsigma _{GL}(0)/(\pi s)]^2$ for our samples, where m and M are the effective mass of the quasi-particles along the layers plane and perpendicular to the layers respectively, $\varsigma _{GL}(0)^2= \phi_{0}/(2\pi T_c |dH_{c2}/dT|$ is the Ginszburg-Landau coherence length at T=0, $\phi_{0}$ is the quantum flux, and $s\sim13 \AA$ \cite{bulk,crystal} is the c-axis lattice constant of the system. The parameter $r$, first defined in Ref. \onlinecite{klemm1}, carries important information about the dimensionality of the system. It is worth mentioning that the above expression for $r$ defined in Ref. \onlinecite{ger}  was obtained for a layered system with the magnetic field applied perpendicular to the layers, and coincides with the expression for $r$ defined by 
Klemm et al. \cite{klemm1} for systems in the dirty limit when calculated at $T=T_c$.  The calculated values are: $r$=0.11 for the crystal with $T_c$=28 K where we used $|dH_{c2}/dT|$=55 kOe/K, $\sqrt{M/m}=3$ and $\varsigma _{GL}(0)=14.6 \AA$, and $r$=0.13 for the crystal with $T_c$=32K where we used $|dH_{c2}/dT|=40 kOe/K$, $\sqrt{M/m}=\gamma=3$ and $\varsigma _{GL}(0)=15.9 \AA$. A plot of the reduced field $H/H{c2}(0)$ vs. $r$ is presented in Figure 9 of Ref. \onlinecite{klemm1}, which helps to identify the dimensionality behavior of fluctuations in a given system when the value of $r$ is know, and also predicts if this system  can exhibit a field-induced-dimensionality crossover (3D toward 2D) in the vicinity of $T_c$. For our samples with $r\sim 0.1$ a D=3 is expected, which agrees with the 3D-LLL scaling analysis results, but a dimensionality crossover (3D to 2D) is predicted to occur  for fields higher than $0.5 H_{c2}(0)$. Since the system has an anisotropy parameter lying between 2 and 3, we also calculate the value of $r $ for  $\sqrt{M/m}=2$, which produced $r \sim 0.3$ for both samples, corresponding to a dimensionality D=3 with no predicted dimensionality crossover induced by field. Values of $T_c(H)$ obtained for both samples from the linear extrapolation of the reversible magnetization and from the 3D-LLL scaling are plotted with values of Tirr, and  shown in the inset of Fig. 4b. The resulting values of $dH_{c2}/dT$ for each sample are shown in this figure and are
in reasonable agreement with values presented in the literature \cite{crystal,welp}. 

Finally, we perform an analysis of the asymptotic behavior of the order parameter amplitude near the onset of superconductivity. The approach used is based on  the conventional theory of the upper critical field $H_{c2}$ \cite{abrikosov}, where the magnetic induction $B$ obtained from the Ginzburg-Landau equation can be expressed as \cite{deGennes}
\begin{equation} 
B=H-\frac{4\pi e \hbar}{mc}|\psi|^2 ,
\end{equation}
where $\psi$ is the superconducting order parameter. The magnetization $M=(B-H)/4\pi$ is then given by
\begin{equation} 
M=-\frac{e \hbar}{mc}|\psi|^2 .
\end{equation}
Within the Abrikosov approximation \cite{abrikosov}, it follows that $\sqrt{M}$ is directly proportional to the average amplitude of the order parameter.  For the magnetization data, the above equation is valid for the entire reversible region, where we applied the linear extrapolation method  to estimate $T_c(H)$. Near the superconducting transition the temperature dependence of the magnetization can be expressed as $\sqrt{M} \propto [T_c(H)-T]^m$, in terms of the mean-field transition temperature $T_c(H)$.  The mean-field exponent is given by $m=1/2$ for a $s$-wave BCS superconductors \cite{deGennes}. This analysis can infer the existence of phase fluctuations in the order parameter \cite{kwon,alvarenga}.  
\begin{figure}[t]
\includegraphics[width=\linewidth]{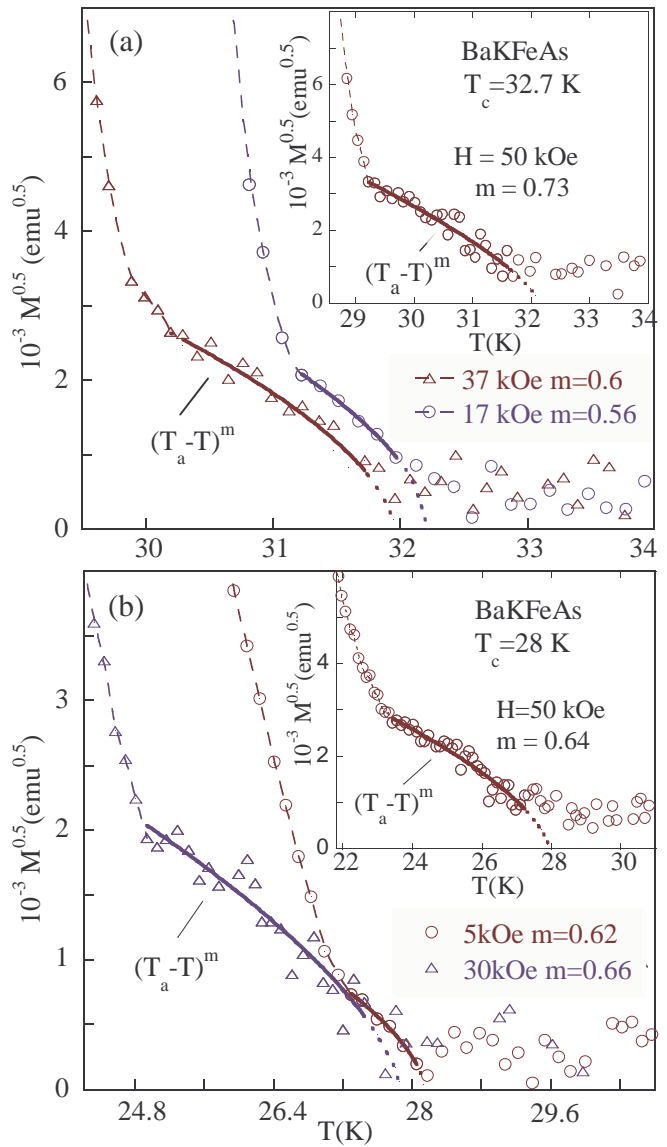}
\caption{Selected isofield curves of $\sqrt{M}$ vs.\ $T$ where $M$ is the reversible magnetization. (a) sample with $T_c$=32.7~K. Inset: curve for H= 50 kOe.  (b) sample with $T_c$=28~K. Inset: curve for H=50 kOe. } \label{fig5}
\end{figure}

Figure~5a and 5b and their insets show selected curves of $\sqrt{M}vs.T$ where it is possible to see that the asymptotic behavior of the reversible region is quite distinct, showing the shape of the gap (opening) in the vicinity of $T_c(H)$. This  clearly indicates the region (temperature window) where  the analysis should be performed.  The solid lines represent the fitting of the selected data to a form  $\sqrt{M} \propto [T_a(H)-T]^m$, where $T_a(H)$ denotes the apparent transition temperature, and $m$ is the fitting exponent. Resulting values of $T_a(H)$ are plotted in the inset of Fig. 4b. It should be mentioned that the values of $T_a(H)$ do not follow the trend suggested in the main figures, with $dT_a(H)/dH<0$ as followed by $T_c(H)$. The latter is exemplified by the curves shown in each inset, where the value of $T_a(H)$  for a higher field curve is little higher than the value for a lower field. This prevented all analyzed data to be shown in one figure, since the curves intercept each other. Results of  these analysis show that values of $T_a(H)$ (dotted lines show extrapolation of the fittings to $T_a(H)$) are slightly larger than values of $T_c(H)$ as well values of the exponent $m$ are larger than the expected value $1/2$. These deviations might be associated to the spread of the data, but we also note that values of $m$ are consistently larger than $1/2$ and increases with field for both samples, which might be not casual but an effect of  phase fluctuations. This can be due to the existence of nodes in the order parameter of this system.  It is shown in Ref. \onlinecite{kwon} that phase fluctuations play an important role when the order parameter has a node,  with an overall effect that reduces the density of states changing the shape of the gap in the vicinity of $T_c$. As a result,  the value of the exponent $m$ increases above the expected value $1/2$ \cite{alvarenga} as observed here. Within this scenario, $T_a(H)$ represents the onset of phase coherence, which occurs a little above the mean field temperatures $T_c(H)$. Since magnetic field enhances fluctuations effects in layered systems \cite{ullah} the exponent $m$ is expected to increase with field, as observed. 
It should be mentioned that the existence of nodes in the order parameter would produce an angular anisotropy in Arpes measurements which was not observed in Refs. \onlinecite{epl} and \onlinecite{germany}. On the other hand the autors of Ref. \onlinecite{mazin} developed an extended s-wave fully gaped theory for iron-pnictides with unconventional pairing which shows that the above mentioned angular anisotropy of the gap due to nodes would not be observed in the case that the gap change sign between the different Fermi Surfaces. More recently a theoretical work also developed assuming unconventional pairing for these systems explains why nodes appears in some experiments and do not in others \cite{scalapino}. 

As a final remark it is important to notice that all curves of Figs~5a and 5b clearly show a gap closing at $T_a(H) \simeq T_c(H)$  which suggests that our results follow the behavior expected for a conventional multi-band s-wave symmetry superconductor \cite{cvetepl,stanevPRB78, kivelson}.  On the other hand, since we observe a value of the exponent $m$ larger than $1/2$ which might be related to the existence of nodes in  the gap, the possibility that the pairing is unconventional as proposed in Refs. \onlinecite{mazin} and \onlinecite{nodes2} can not be discarded.

In conclusion, we observe three-dimensional fluctuations magnetization of the lowest-Landau-level type in  $Ba_{1-x}K_xFe_2As_2$ in a temperature window exceeding 3 K above $T_c(H)$. Analysis of the reversible magnetization allowed a study of the asymptotic behavior of the gap near $T_c(H)$.  Results of the analysis produced values for the exponent $m$ larger than $1/2$ suggesting that phase fluctuations are important near $T_c$ in  $Ba_{1-x}K_xFe_2As_2$ which is consistent with nodes in the gap. 
\begin{acknowledgements}
LG and ADA thanks support from the Brazilian agencies CNPq and FAPERJ. 
\end{acknowledgements}

%%%%%%%%%%%%%%%%%%%%%%%%%%%%%%%%%%%%%%%%%%%%%%%%%%%%

\begin{thebibliography}{99}
\bibitem{japao}Y. Kamihara, T. Watanabe, M. Hirano, and H. Hosono, J. Am. Chem. Soc. {\bf 130}, 3296 (2008).
\bibitem{milton}M.S. Torikachvilli , N. Ni, S.L. BudÕko and P.C. Canfield, Phys. Rev. Lett. {\bf 101}, 057006 (2008).
\bibitem{twogaps}C. Ren, Z.S. Wang, H.Q. Luo, H. Yang, L. Shan, Hai-Hu Wen, Phys. Rev. Lett. 101, 257006 (2008).
\bibitem{0808.2392}Z.S. Wang, H.Q. Luo, C. Ren, Hai-Hu Wen, Phys. Rev. B78, 140501(R) (2008).
\bibitem{0807.3137} H.Q. Yuan, J. Singleton, F.F. Balakirev, S.A. Baily, G.F. Chen, J.L. Luo and N.L. Wang, Nature {\bf 457}, 565 (2009)
\bibitem{CvLuo}G. Mu, H. Luo, Z-S Wang, L. Shan, C. Ren and Hai-Hu Wen, arXiv:cond-mat/0808.2941v1
\bibitem{0812.1188}G. Mu, H.Q. Luo, Z. Wang, L. Shan, C. Ren and Hai-Hu Wen, Phys. Rev. B {\bf 79}, 174501 (2009)
\bibitem{welp}U Welp, R. Xie, A.E. Koshelev, W.K. Kwok, H.Q. Luo, Z.S. Wang, G. Mu and Hai-Hu Wen, Phys. Rev. B {\bf 79}, 094505 (2009)
\bibitem{0807.4488}M. M. Altarawneh, K. Collar, and C. H. Mielke,N. Ni, S. L. Bud'ko, and P. C. Canfield, Phys. Rev. B 78, 220505(R) (2008)
\bibitem{0901.0177}H. Fukazawa, T. Yamazaki, K. Kondo, Y. Kohori, N. Takeshita, P.M. Shirage, K Kihou, K. Miyazawa, H. Kito, H. Eisaki, A. Iyo, J. Phys. Soc. Jpn. {\bf 78}, 033704 (2009)
\bibitem{0811.4668}J.G. Checkelsky, Lu Li, G.F. Chen, J.L. Luo, N.L. Wang and N.P. Ong, arXiv:cond-mat/0811.4668
\bibitem{epl}H. Ding, P. Richard, K. Nakayama, K. Sugawara, T. Arakane, Y. Sekiba, A. Takayama, S. Souma, T. Sato, T. Takahashi, Z. Wang, X. Dai, Z. Fang, G.F. Chen, J.L. Luo and N.L. Wang, Euro. Phys. Lett {\bf 83} 47001 (2008)
\bibitem{germany}D.V. Evtushinsky, D.S. Inosov, V.B. Zabolotnyy, M.S. Viazovska, R. Khasanov, A. Amato, H.-H. Klauss, H. Luetkens, Ch. Niedermayer, G.L. Sun, V. Hinkov, C.T. Lin, A. Varykhalov, A. Koitzsch, M. Knupfer, B. Bchner, A.A. Kordyuk, and S.V. Borisenko, New J. Phys. {\bf 11}, 055069 (2009)
\bibitem{chia}Elbert E. M. Chia, D. Talbayev, J. X. Zhu, H. Q. Yuan, T. Park, J. D. Thompson, G. F. Chen, J. L. Luo, N. L. Wang and A. J. Taylor, arXiv:cond-mat/0809.4097
\bibitem{0807.3786}H. Yang, H.Q. Luo, Z.S. Wang, and Hai-Hu Wen, Appl. Phys. Lett. {\bf 93}, 142506 (2008).
\bibitem{germany2}J.T. Park, D.S. Inosov, Ch. Niedermayer G. L. Sun, D. Haug, N.B. Christensen, R. Dinnebier, A.V. Boris, A.J. Drew, L. Schulz, T. Shapoval, U. Wolff, V. Neu, Xiaoping Yang, C.T. Lin, B. Keimer, and V. Hinkov, Phys. Rev. Lett {\bf102}, 117006 (2009).
\bibitem{klemm1}R.A. Klemm, M.R. Beasley, and A. Luther, Phys. Rev. B{\bf 8}, 5072 (1973).
\bibitem{klemm2}W.C. Lee, R.A. Klemm, and D.C. Johnston, Phys. Rev. Lett. {\bf 63}1012 (1989).
\bibitem{tinkham}M. Tinkham, {\it Introduction to Superconductivity}, (McGraw-Hill Inc., 2nd edition, New York, 1996)
\bibitem{ger}R.R. Gerhardts, Phys. Rev. B{\bf9}, 2945 (1974).
\bibitem{ullah}S. Ullah and A. T. Dorsey, Phys. Rev. Lett. {\bf 65}, 2066 (1990)
\bibitem{tesa}Z. Tesanovic, L. Xing, L. Bulaevskii, Q. Li, and M. Suenaga, Phys. Rev. Lett. {\bf 69}, 3563 (1992).
\bibitem{rosenstein}B. Rosenstein, B.Y. Shapiro, R. Prozorov, A. Shaulov and Y. Yeshurun, Phys. Rev. B{\bf 63}, 134501 (2001)
\bibitem{kwon}H-J Kwon, Phys. Rev. B{\bf 59}, 13600, (1999)
\bibitem{beck}P. Curty and H. Beck, Phys. Rev. Lett. {\bf91}, 257002, (2003)
\bibitem{jesus}L. Cabo, J. Mosqueira, and F. Vidal, Phys. Rev. Lett. {\bf 98}, 119701 (2007)
\bibitem{nb}S. Salem-Sugui Jr., M. Friesen, A. D. Alvarenga, F. G. Gandra, M. M. Doria and O. F. Schilling, Phys. Rev. B{\bf66}, 134521 (2002)
\bibitem{said}S. Salem-Sugui Jr., A. D. Alvarenga, V.N. Vieira and O.F. Schilling, Phys. Rev. B {\bf 73}, 012509 (2006).
\bibitem{paulo}R. Menegotto Costa, P. Pureur, L. Ghivelder, J. A. Camp‡, and I. Rasines, Phys. Rev. B 56, 10836 (1997)
\bibitem{alvarenga}S. Salem-Sugui Jr and A.D. Alvarenga, Phys. Rev. B {\bf 77},104533 (2008)
\bibitem{bulk}M. Rotter, M. Tegel, and D. Johrendt, Phys. Rev. Lett. {\bf} 101, 107006 (2008).
\bibitem{crystal}H.Q. Luo, Z.S. Wang, H. Yang, P. Cheng, X. Zu, and Hai-Hu Wen, Supercond. Sci. Technol. {\bf 21}, 125014 (2008)
\bibitem{0807.3950}H. Chen, Y. Ren, Y. Qiu,, Wei Bao, R.H. Liu, G. Wu, T. Wu, Y.L. Xie, X.F. Wang, Q. Huang and X.H. Chen, Euro. Phys. Lett. {\bf 85} 17006 (2009)
\bibitem{norman}M. Norman, Physics {\bf 1}, 21 (2008)
\bibitem{nodes2}V. Mishra, G. Boyd, S. Graser, T. Maier, P.J. Hirschfeld, D.J. Scalapino, Phys. Rev. B {\bf 79}, 094512 (2009)
\bibitem{stanevPRB78}V. Stanev, J. Kang, and Z. Tesanovic, Phys. Rev. B {\bf 78} 184509, (2008)
\bibitem{cvetepl}C. Cvetkovic and Z. Tesanovic, Europhysics Lett. {\bf 85}, 37002 (2009)
\bibitem{mazin}I.I. Mazin, D.J. Singh, M.D. Johannes, and M.H. Du, Phys. Rev. Lett. {\bf 101}, 057003, (2008)
\bibitem{welp2}U. Welp, S. Fleshler, W.K. Kwok, R.A. Klemm, V.M. Vinokur, J. Downey, B. Veal and G.W. Crabtree, Phys. Rev. Lett. {\bf 67}, 3180 (1991)
\bibitem{zac}S. Salem-Sugui, Jr., and E. Z. da Silva, Physica C {\bf 235},1919 (1994)
\bibitem{timusk}T. Timusk and B. Statt, Rep. Prog. Phys. {\bf 62}, 61 (1999)
\bibitem{kpaper}S. Salem-Sugui Jr., M.M. Doria, A.D. Alvarenga, V.N. Vieira, P.F. Farinas, and J.P. Sinnecker, Phys. Rev. B {\bf 76}, 132502  (2007)
\bibitem{mauro}M.M. Doria and S. Salem-Sugui Jr., Phys. Rev. B 78, 134527 (2008)
\bibitem{chennature}T. Y. Chen, Z. Tesanovic, R. H. Liu, X. H. Chen, and C. L. Chien, Nature {\bf 453} 1224 (2008)
\bibitem{abrikosov}A. Abrikosov, Zh. Éksp. Teor. Fiz. {\bf 32}, 1442 (1975) (Sov. Phys. JETP {\bf 5}, 1174 (1957)
\bibitem{deGennes}P.G. deGennes, Superconductitivity of Metals and Alloys,(1989)
\bibitem{scalapino}T.A. Maier, S. Graser, D.J. Scalapino, and P.J. Hirschfeld, Phys. Rev. B {\bf 79}, 224510 (2009)
\bibitem{kivelson}S.A. Kivelson and H. Yao, Nature Materials {\bf 7}, 927 (2008)
\end{thebibliography}
\end{document}